\documentclass{emulateapj}    

\newcommand{\gtapprox}{\raisebox{-0.5ex}{$\,\stackrel{>}{\scriptstyle
\sim}\,$}}
\newcommand{\ltapprox}{\raisebox{-0.5ex}{$\,\stackrel{<}{\scriptstyle
\sim}\,$}}

\newcommand{\rw}{r_{\rm w}}

\newcommand{\Mdota}{\dot M_{\rm a}}
\newcommand{\Mdotw}{\dot M_{\rm w}}

\newcommand{\Tb}{T_{\rm b}}
\newcommand{\Te}{T_{\rm e}}
\newcommand{\Ne}{N_{\rm e}}

\shorttitle{Radio Core Emission from Radio-Quiet Quasars}
\shortauthors{Blundell \& Kuncic}

\begin{document}


\title{On the Origin of Radio Core Emission in Radio-Quiet Quasars}


\author{Katherine M.\ Blundell}
\affil{Department of Physics, University of Oxford, Keble Road, Oxford OX1 3RH, UK}
\email{k.blundell@physics.ox.ac.uk}

\and

\author{Zdenka Kuncic}
\affil{School of Physics, University of Sydney, Physics Road, Sydney,
  NSW 2006, Australia} 
\email{z.kuncic@physics.usyd.edu.au}

\begin{abstract}
We present a model for the radio emission from radio-quiet quasar
nuclei.  We show that a thermal origin for the high brightness
temperature, flat spectrum point sources (known as radio ``cores'') is
possible provided the emitting region is hot and optically-thin.  We
hence demonstrate that optically-thin bremsstrahlung from a slow,
dense disk wind can make a significant contribution to the observed
levels of radio core emission.  This is a much more satisfactory
explanation, particularly for sources where there is no evidence of a
jet, than a sequence of self-absorbed synchrotron components
which collectively conspire to give a flat spectrum.  Furthermore,
such core phenomena are already observed directly via milli-arcsecond
radio imaging of the Galactic microquasar SS\,433 and the active
galaxy NGC\,1068.  We contend that radio-emitting disk winds must be
operating at some level in radio-loud quasars and radio galaxies as
well (although in these cases, observations of the radio cores are
frequently contaminated/dominated by synchrotron emission from jet
knots).  This interpretation of radio core emission mandates mass
accretion rates that are substantially higher than Eddington.
Moreover, acknowledgment of this mass-loss mechanism as an AGN
feedback process has important implications for the input of energy
and hot gas into the inter-galactic medium (IGM) since it is
considerably {\it less directional} than that from jets.

\end{abstract}

\section{Introduction}

Radio-quiet quasars are strongly accreting yet lack the extensive
100-kpc scale jets that steadily transport away energy and angular
momentum from the nuclei of radio-loud quasars and radio galaxies.
The weak radio core emission associated with radio-quiet quasars has
been revealed by a succession of very long baseline interferometry
(VLBI) observations usually to be compact on milliarcsecond ($\sim$parsec) size scales \citep{Blundell96,Falcke96,BlundBeas98,Ulvestad05}.
The luminosities of the cores in such objects can be as high as VLBI
measurements of cores in many radio-loud quasars ($10^{21}$ --
$10^{24}\,{\rm W\,Hz^{-1}\,sr^{-1}}$) when imaged with resolution of a
few milli-arcsec; comparable brightness temperatures ($10^6$ --
$10^9$\,K) are revealed by these observations as well.  The physical
size-scales of the observed cores reported in these papers for
radio-loud and radio-quiet quasars have a robust upper limit of a few
cubic parsec even at redshift $z \sim 1$.

Observations of some nearby Seyfert nuclei reveal a flat-spectrum
component on milli-arcsecond scales with a misalignment of up to
$90^\circ$ from the linear radio structure on arcsecond scales
\citep[e.g.\,][]{Middelberg04}.  In some cases, a non-thermal origin for radio
cores is directly ruled out by observations: VLBI observations of the
nucleus in NGC\,1068 in particular reveal linear parsec-scale radio
structure perpendicular to the radio jet axis \citep*{GBO97}, 
with a brightness temperature, $\Tb \approx 10^6$\,K, too low to
be consistent with Synchrotron Self-Absorption (SSA) 
\citep{GBO96}. Its radio spectrum measured
between $5$\,GHz and $8.4$\,GHz is $\propto \nu^{0.3}$.  \citet{GBO04}
suggest that the observed radio properties of the NGC\,1068 nucleus
are best explained in terms of free-free emission from a disk wind.

The purpose of this paper is to construct a model for AGN cores based
on this and on observations of the resolved disk-wind outflow from
the nuclear regions of the Galactic microquasar SS\,433, which arises
from optically-thin bremsstrahlung \citep{Blundell01}, as an
alternative to the long-standing model of a sequence of synchrotron
self-absorbed components which conspire to give a smooth flat
spectrum \citep{Cotton80}.

\subsection{Beyond cosmic conspiracy theory}

The interpretation of the ``cosmic conspiracy'' core model --- that is, a
sequence of distinct homogeneous components each having a peaked
spectrum which together give rise to the integrated smooth flat
spectrum core as in the case of the BL Lacertae object 0735+178
\citep{Cotton80} --- has been applied to many observations of cores in the
last three decades.    In many subsequent observations of many
different objects there has been no direct observational evidence
supporting this paradigm and indirect evidence against it,  
even under the scrutiny of sub-milliarcsec,
multi-epoch observations \citep[e.g.\,][]{Rantakyro98,Gomez00,Jorstad05,Ly07}.
There is simply no evidence in support of multiple self-absorbed
components which conveniently have successively higher turn-over
frequencies to explain the overall integrated spectral properties of stationary cores.
We are compelled, therefore, to explore the validity of an alternative
model in which the mechanism responsible for the radio
emission is optically-thin bremsstrahlung arising from a disk wind,
rather than embryonic, and moving, jet knots.   

\section{{Disk Winds as a Source of Flat Spectrum Radio
Core Emission?}}

A common misconception is that high observed brightness temperatures
necessarily imply a non-thermal source. Whilst it is true that a high
brightness temperature $\Tb$ rules out optically-thick thermal emission 
(because $\Tb$ cannot
exceed the gas temperature, $\Te$, which is low in the case of an
optically thick source because a blackbody is an efficient radiator),
optically-\emph{thin} thermal emission cannot be ruled out as
readily. The brightness temperature of an optically-thin thermal
source of size $R$ is $\Tb \approx c^2 j_\nu R/(2k\nu^2)$, where
$j_\nu$ is the volume emissivity at frequency $\nu$. At temperatures above $\Te \simeq
10^7\,{\rm K}$, gas is completely ionized and if the plasma is
sufficiently dense and remains subrelativistic, then two-body (i.e.\ free-free) processes dominate
Compton processes. The brightness temperature can then be expressed as
$\Tb \approx \tau_\nu^{\rm ff} \Te$, where $\tau_\nu^{\rm ff}
\ltapprox 1$ is the free-free optical depth. Thus, a high brightness
temperature can arise from a thermal plasma provided it is hot and
marginally optically thin.  Moreover, because the spectral index of
bremsstrahlung emission is $\alpha \approx - (h\nu / k\Te) \log e$
(where $\alpha \equiv d \log S_\nu / d \log \nu$ and $S_{\nu}$ is the flux density at frequency $\nu$), a flat spectrum is
naturally produced at radio frequencies ($h\nu \ll k\Te$).

In the following, we consider a thermal accretion disk wind that is
present at radii $r \gtapprox \rw$, where $\rw$ is the disk radius at which
the wind is launched \citep*[see][]{BegMcKSh83,KingPounds03}. Within
this radius, outflows are likely to become Poynting-flux-dominated and
form a magnetized corona and/or jet. Self-absorbed synchrotron
radiation from the jet can of course contribute to the unresolved core radio
emission. We consider the possibility that the thermal disk wind also
contributes to flat-spectrum core radio emission through
optically-thin free-free radiation.  At the base of the wind, the
electron number density, $\Ne$, can be related to the mass outflow
rate, $\Mdotw$, via the continuity equation
\begin{equation}
\Ne (\rw) \simeq \frac{\Mdotw}{4\pi f_\Omega {\rw}^2 \mu m_{\rm p}
  v_{\rm w}} \qquad ,
\label{e:Ne1}
\end{equation}
where $v_{\rm w}$ is the escape speed at this radius 
and $f_\Omega = \Omega /4\pi \ll 1$ is the geometrical covering 
factor of the outflow. This gives
\begin{eqnarray}
\Ne \simeq 1 &\times& 10^{12} \, f_{\Omega ,0.1}^{-1} \,
\frac{\Mdotw}{M_\odot \, {\rm yr}^{-1}}
\left( \frac{\rw}{10^{15}\,{\rm cm}} \right)^{-2} \nonumber \\
&\times& \left( \frac{v_{\rm w}}{500\,{\rm km\,s}^{-1}} \right)^{-1}
{\rm cm}^{-3}
\label{e:Ne2}
\end{eqnarray}
where $f_{\Omega ,0.1} = f_\Omega /0.1$ and where we have used a mean
molecular weight $\mu =0.5$ for fully ionized hydrogen.  At these
densities, the base of the wind is highly opaque
\citep[see][]{KingPounds03}.  We now consider the properties of the
wind further out.

Optically-thin thermal emission from the disk wind can become
important beyond a photospheric radius $r_{\rm ph}$ where the wind
becomes transparent.  This is the radius at which the effective
optical depth, $\tau_{\rm eff} = \sqrt{\tau_{\rm ff} ( \tau_{\rm ff} +
\tau_{\rm es})}$ \citep{RybLight79}, is unity as viewed by a distant
observer. Here, $\tau_{\rm ff}$ is the optical depth due to free-free
absorption and $\tau_{\rm es}$ is the electron scattering optical
depth. To determine $r_{\rm ph}$ from the condition $\tau_{\rm eff}
\approx 1$, we first calculate separately the scattering and
absorption photospheric radii $r_{\rm es}$ and $r_{\rm ff}$ and
determine the physical conditions under which they are approximately
equal. Using $\Ne \propto r^{-2}$ from mass continuity
(eqn\,\ref{e:Ne1}), the radius $r_{\rm es}$ at which $\tau_{\rm es}
\approx 1$ is determined from $\int_{r_{\rm es}}^\infty \sigma_{\rm T}
\Ne \, dr \approx 1$, where $\sigma_{\rm T}$ is the Thomson scattering
cross section. This gives
\begin{eqnarray}
r_{\rm es} &\approx& \frac{\sigma_{\rm T} \Mdotw}{4\pi f_\Omega \mu
   m_{\rm p} v_{\rm w}} \\
&\approx& 8 \times 10^{17} \, f_{\Omega ,0.1}^{-1} \,
\frac{\Mdotw}{M_\odot \, {\rm yr}^{-1}}
\left( \frac{v_{\rm w}}{500\,{\rm km\,s}^{-1}} \right)^{-1}
\, {\rm cm}. \nonumber
\end{eqnarray}
Similarly, the radius $r_{\rm ff}$ at which $\tau_{\rm ff} \approx 1$
is determined from $\int_{r_{\rm ff}}^\infty
\kappa_\nu^{\rm ff} \, dr \approx 1$, where $\kappa_\nu^{\rm ff} \approx
0.018 \, \bar g_{\rm ff} \, \nu^{-2} \Te^{-3/2} \Ne^2 \, {\rm cm}^{-1}$ is 
the free-free
absorption coefficient and $\bar g_{\rm ff}$ is the velocity-averaged 
free-free Gaunt
factor \citep{RybLight79}. This gives
\begin{eqnarray}
\label{e:rff}
r_{\rm ff} &\approx& 0.39 \, \bar g_{10}^{1/3} \, \nu^{-2/3} \,
\Te^{-1/2}
\left( \frac{\Mdotw}{4\pi f_\Omega  \mu m_{\rm p}v_{\rm w}}
\right)^{2/3} \, {\rm cm}
  \\
&\approx& 8 \times 10^{17} \, \bar g_{10}^{1/3} \, \nu_{\rm
8GHz}^{-2/3} \, T_7^{-1/2} \,
f_{\Omega ,0.1}^{-2/3}
\nonumber \\
&\times& \left( \frac{\Mdotw}{M_\odot \, {\rm yr}^{-1}}  \right)^{2/3}
\left( \frac{v_{\rm w}}{500\,{\rm km\,s}^{-1}} \right)^{-2/3} \, {\rm
cm}
\nonumber
\end{eqnarray}
where $\nu_{\rm 8GHz} = \nu / 8\,{\rm GHz}$, $T_7 = \Te/10^7\,{\rm
K}$ and $\bar g_{10} = \bar g_{\rm ff}/10$.
Thus, a hot disk wind that is optically-thin to both electron
scattering and radio-frequency free-free absorption
can exist beyond a photospheric radius
$r_{\rm ph} \simeq 0.1 - 1 \,{\rm pc}$.

We now consider the bremsstrahlung radio power emitted by the
optically-thin part of the wind.
The specific bremsstrahlung luminosity is
\begin{equation}
L_\nu \simeq  4\pi f_\Omega  \int_{r_{\rm ph}}^{\infty}  j_\nu^{\rm
ff}  \, r^2 \, dr
\qquad ,
\label{e:Lnuff}
\end{equation}
where $j_\nu^{\rm ff}$ is the free-free volume emissivity.
For an electrically neutral hydrogen plasma,
$j_\nu^{\rm ff} \simeq 6.8 \times 10^{-37} \, \bar g_{10} \, T_{\rm
e}^{-1/2} N_{\rm e}^2
\exp (-h\nu/kT_{\rm e})
     \,   {\rm erg \, s^{-1} \, cm^{-3} \, Hz^{-1} }$
      \citep{RybLight79}.
This gives a specific luminosity per solid angle
\begin{eqnarray}
\label{e:Lnu}
L_{\nu ,\Omega} &=& \frac{L_\nu}{\Omega} \simeq 2 \times 10^{-36} \,
\, \bar g_{10} \,T_{\rm e}^{-1/2}
\left( \frac{\Mdotw}{4\pi f_\Omega  \mu m_{\rm p}v_{\rm w}}
\right)^{2} \, r_{\rm ph}^{-1}
\nonumber \\
&& \hspace{2.0truecm} {\rm erg \, s^{-1} Hz^{-1} sr^{-1}} \nonumber \\
&\approx& 3 \times 10^{20} \, \, \bar g_{10} \,T_7^{-1/2} \,
f_{\Omega ,0.1}^{-2} \,
\left( \frac{\Mdotw}{M_\odot \, {\rm yr}^{-1}}  \right)^2 \\
&\times& \left( \frac{v_{\rm w}}{500\,{\rm km\,s}^{-1}}
\right)^{-2}
\left( \frac{r_{\rm ph}}{0.1 \,{\rm pc}} \right)^{-1}
\, {\rm W \, Hz^{-1} \, sr^{-1}}.
\nonumber
\end{eqnarray}

\subsection{{Observed core luminosities}}

\citet{White07} applied an innovative technique to measure radio core
emission from the quasars in the SDSS DR3 quasar catalog \citep{Schneider05}. By
stacking images from the FIRST survey \citep{Becker95}, they found
radio core luminosities in the range of $10^{21}$ -- $10^{24} {\rm W
Hz^{-1} sr^{-1}}$, consistent with luminosities measured on
milli-arcsecond scales
\citep{Blundell96,Falcke96,BlundBeas98,Ulvestad05}.  Of course, one
cannot be sure that the core luminosities measured by \citet{White07} 
are not elevated because of
contamination by synchrotron emission from moving jet knots.  Indeed,
\citet*{Blundell03} have direct observational evidence for a highly
relativistic jet in the bona fide radio-quiet quasar PG\,1407+263.

\section{{Super-Eddington Accretion?}}

A comparison of the radio wind luminosities predicted by
eqn\,(\ref{e:Lnu}) and observed radio core luminosities indicates that
disk winds can make a non-negligible contribution to observed radio
core luminosities only if there is significant mass loss, with
$\Mdotw$ exceeding the Eddington limit, $\dot M_{\rm Edd} \simeq 3\,
\eta_{0.1} M_8 M_\odot \, {\rm yr}^{-1}$, where $M = 10^8 M_8$ is the
black hole mass and $\eta = 0.1
\eta_{0.1}$ is the  radiative efficiency. This in turn implies
accretion rates $\dot M_{\rm a} \gg \dot
M_{\rm Edd}$. The main role of the disk wind is then to remove excess
angular momentum, thereby reducing the accretion rate at smaller radii.

Are super-Eddington accretion and outflow rates plausible?  We remark
that the Eddington limit is based on simplifying (and incorrect)
assumptions of spherical symmetry and a fiducial $10$\% radiative
efficiency.  The brightest radio galaxies and quasars have bolometric
luminosities $L \simeq 10^{46-47}\,{\rm erg \, s}^{-1}$, requiring
mass accretion rates of $\Mdota \gtapprox 1 - 10 \, M_\odot \, {\rm
yr}^{-1}$.  This is a lower limit because the observed radiative
output is almost certainly only a small fraction of the total power
extracted from accretion, much of which can be converted into low
radiative efficiency phenomena such as magnetized jets and/or a corona
\citep[see][]{KunBick04,KunBick07b}.

Similarly, mass outflow rates for warm, dense ionized gas inferred
from X-ray absorption spectra of quasars and Seyferts are typically
$0.1 - 10 \, M_\odot \, {\rm yr}^{-1}$
\citep{Chartas02,Pounds03a,Pounds03b,obrien05,PoundsPage06}. Again,
these values imply super-Eddington accretion. In the case of SS433,
the mass outflow rate inferred for the observed radio wind is $\simeq
4 \times 10^{-4} M_\odot \, {\rm yr}^{-1}$ \citep{Blundell01}. If this
Galactic microquasar is a $10 M_\odot$ black hole, then this implies
an astonishing mass loss rate of $\Mdotw \sim 10^3 \dot M_{\rm Edd}$.

\section{Implications}

\subsection{Correlated radio and optical/UV  emission}
If radio core emission arises at least in part from an optically-thin
disk wind, then we may expect to observe a correlation between the
radio wind emission and the optical/UV blackbody disk emission. This
is expected because the disk wind can modify the local radial
structure 
of the accretion flow, resulting in a disk emission spectrum
that is redder and dimmer than that of a disk without outflows
\citep{KunBick07a}.  Such a correlation has indeed been reported by
\citet[their fig.\ 14]{White07}, who find that the SDSS DR3 quasars with redder
colours tend to have higher mean radio flux densities and are
radio-louder than quasars with standard colours\footnote{A positive
correlation between radio-loudness and bluer quasars is also evident
in the catalog,  although \citet{White07} suggest this is most likely due to beamed
synchrotron emission, i.e.\ jet contamination.}.

Additionally, \citet[their fig.\ 9]{White07} find a remarkably tight correlation
between the $5\,{\rm GHz}$ median radio luminosity and the absolute 
magnitude at $2500$\,\AA \, restframe wavelength, 
with the specific radio and optical luminosities related via $L_R
\sim L_{\rm opt}^{0.85}$.
The disk wind model predicts that $L_R \approx
L_{\rm opt}$ and that optical bremsstrahlung wind emission should begin to
contribute significantly to the observed ($K$-corrected) mean optical
quasar luminosity at $2500$\,\AA
\, (restframe), $L_{\rm
   opt} \simeq 4 \times 10^{30} \, {\rm erg \, s^{-1} \, {\rm
     Hz}^{-1}}$, when $\Mdotw \gtapprox 10 \, M_\odot \, {\rm
   yr}^{-1}$ (see eqn\,\ref{e:Lnu}).
Thus, unless the mass loss rate is extraordinarily high (that is, higher than the rates inferred from other independent observations -- c.f. \S\,3), it is likely  
that the optically-thick disk emission, rather than the optically-thin disk wind emission,  is primarily responsible for the observed $2500$\,\AA \,
emission. Remarkably, when \citet{White07} calculate the radio
loudness parameter, $R^\star = L_R /L_{\rm opt}$, adjusted to remove
the strong radio/optical correlation, they find $\log R^\star \sim 0$ (their fig.\ 11).
While it is not clear that this result is definitive, because of
their assumption that all radio cores at all redshifts have a straight
spectral index of $\alpha = -0.5$ (for which there are many
counterexamples, some of them seemingly systematic with redshift), we
note that our disk wind model provides a natural explanation for $\log R^\star$
(corrected for blackbody disk emission) being consistently close to zero.

The parsec-scale cores of radio-quiet quasars are comparable in
luminosity, size and brightness temperature to those in radio-loud
quasars.  We suggest that the model explored in this paper has
applicability to the stationary, unresolved, flat spectrum components
revealed with milli-arcsecond VLBI observations, now an established
technique for revealing moving and evolving synchrotron-emitting jet
knots \citep[e.g.\,][]{Gomez00,Jorstad05,Ly07}, and even revealed with
global VLBI techniques with 10s of micro-arcsecond resolution
\citep{Rantakyro98}.

\subsection{{Duty cycles of quasars and two distinct modes of energy feedback: jets and disk winds}}
\label{sec:rqq}

\citet{BlundellRawlings2000,BlundellRawlings1999} have presented
evidence (from physical arguments and from the observed near-absence
of any relic radio lobes) that lobe emission must disappear on
relatively rapid timescales.  From the similarity between the
duty-cycle of intermittent jet activity in microquasars (their flaring mode) 
and the fraction of quasars that are observed to be radio-loud,
\citet*{Nipoti05} posit that radio-loudness is a function of the epoch
at which a quasar is observed.  The present paper suggests that when
there is no evidence for kpc-scale jets or their lobes, that disappear
rapidly following too much expansion, the observed persistence of weak
unresolved radio cores is a strong indicator that mass-loss via a disk
wind accompanies on-going disk accretion.  Indeed, it seems increasingly likely
that radio-loudness in quasars is a somewhat short-lived phase of
enhanced angular momentum loss via jets, analogous to that in
microquasars, and that the more common means of angular momentum loss
is via longer-lived phases of mass-loss via disk winds.
Separately from the flaring mode associated with jet-ejection episodes in microquasars, \citet{Nipoti05} identify a second, chief mode of energy loss, which they term the ``coupled mode" during which microquasars are observed to have closely coupled X-ray and radio luminosities and to be unresolved ``cores". If the quiescent radio core emission is due to a hot ($\sim 10^7$\,K) disk wind, then  the observed radio/X-ray correlation could be attributed to a common disk origin for the radio (disk wind) and soft X-ray (disk wind + disk blackbody) emission.  It would be interesting to pursue this analogy further in the context of core emission in quasars and AGN (although in this case, only the disk wind, and not the disk blackbody emission, would contribute to the soft X-ray emission) and indeed the extent to which the disk wind, if equatorial as in SS433, plays the role of the putative obscuring torus.

Although it is now widely acknowledged that feedback from AGN is required
to reconcile simulations of galaxy formation with observations, the
details of these processes remain to be explored \citep[e.g.\,][]{Nesvadba06}.
In particular, although jets are invoked as a means of relocating mass
and energy away from the active galactic nucleus, a limitation of this
picture is that jets are by their very nature {\em highly
directional} and relatively light.  Winds from accretion disks, however, are
considerably less directional and heavier and may provide a better means of
dispersing mass/energy into the IGM and explaining the observed
link between growth of supermassive black holes and their host galaxy
properties \citep{King05}.  A mass outflow rate ($\Mdotw 
\gtapprox \dot M_{\rm Edd}$) of hot ($\sim 10^7$\,K) gas may be sufficient to offset
cooling in galaxies.

\section{Concluding Remarks}
We have proposed that the radio cores of radio-quiet quasars are
thermal in origin, arising from an accretion disk wind. This wind
naturally produces flat spectrum radio emission via optically-thin
bremsstrahlung radiation.  The remarkable similarities in radio core
properties of radio-quiet and (many) radio-loud quasars suggests
that a radio-emitting disk wind is present at some level in all
quasars.  The observed luminosities of radio core emission from
quasars implies that they are accreting at super-Eddington rates and
that the disk wind expels most of this matter well before it reaches
the inner accretion flow, thereby providing an efficient mechanism for
angular momentum transport and AGN feedback.

\acknowledgments

We would like to thank the referee for helpful comments on the manuscript.
KMB thanks the Royal Society for a University Research Fellowship and
the Leverhulme Trust for their support.  ZK acknowledges an Australian
Academy of Science Scientific Visits to Europe grant.  
Both authors thank the 1851 Royal Commission for the
support of their research careers, especially most recently for the
reception at Buckingham Palace, London, where this work was conceived.

\end{document}